\documentclass{article}

\usepackage[final,nonatbib]{neurips_2021_ml4ps}

\usepackage[utf8]{inputenc} 
\usepackage[T1]{fontenc}    
\usepackage{hyperref}       
\usepackage{url}            
\usepackage{booktabs}       
\usepackage{amsfonts}       
\usepackage{nicefrac}       
\usepackage{microtype}      
\usepackage{xcolor}         

\usepackage[pdftex]{graphicx}

\title{Simulation-Based Inference of Strong Gravitational Lensing Parameters}

\author{%
  Ronan Legin$^*$$\mathbf{^1}$
  \And
  Yashar Hezaveh$\mathbf{^{1,2}}$
  \And
  Laurence Perreault Levasseur$\mathbf{^{1,2,3}}$
  \And
  Benjamin Wandelt$\mathbf{^{2,4}}$
  \AND
    $^1${\normalfont Universit\'{e} de Montréal} \\
    $^2${\normalfont Flatiron Institute} \\
    $^3${\normalfont Mila} \\    
    $^4${\normalfont Sorbonne Universit\'{e}}\\
    $^*$\texttt{ronan.legin@umontreal.ca}
}

\begin{document}

\maketitle

\begin{abstract}
In the coming years, a new generation of sky surveys, in particular, Euclid Space Telescope (2022), and the Rubin Observatory's Legacy Survey of Space and Time (LSST, 2023) will discover more than 200,000 new strong gravitational lenses, which represents an increase of more than two orders of magnitude compared to currently known sample sizes \cite{2015ApJ...811...20C}. Accurate and fast analysis of such large volumes of data under a statistical framework is therefore crucial for all sciences enabled by strong lensing. Here, we report on the application of simulation-based inference methods, in particular, density estimation techniques, to the predictions of the set of parameters of strong lensing systems from neural networks. This allows us to explicitly impose desired priors on lensing parameters, while guaranteeing convergence to the optimal posterior in the limit of perfect performance.
\end{abstract}

\section{Introduction}

Strong gravitational lensing is a phenomenon in which the light rays of distant galaxies are deflected by the gravity of foreground matter, resulting in the production of multiple images. It is a powerful probe that can map the inner distribution of matter in individual lens galaxies to reveal invaluable information about the physics of the dark matter particle \cite{2016ApJ...823...37H}. It can also provide precise estimates of the expansion rate of the Universe (the Hubble constant), an important measurement, given what is referred to as a current "crisis in cosmology" (a significant tension between the measurements of this parameter by multiple probes) \cite{2020MNRAS.498.1420W}. Strong lensing can also be used to study the magnified images of background galaxies, which are typically some of the most distant galaxies of the Universe, effectively working as a natural telescope \cite{2013ApJ...762...32C}. 

In recent years, convolutional neural networks have been shown capable of providing accurate point estimates of the parameters describing these strong lensing systems more than 10 million times faster than traditional methods \cite{2017Natur.548..555H}. 
Subsequent works have expanded this result to also obtain uncertainty estimates for the predictions made by neural networks. This has been done using approximate Bayesian neural networks trained with variational inference \cite{2017ApJ...850L...7P}. Despite their success at producing accurate measurements in controlled experiments, these procedures involve many levels of approximations (e.g., the choice of the variational distributions for the outputs and the network weights), which cannot be easily quantified or controlled, possibly resulting in biased estimates. Additionally, despite their name, they do not offer a truly Bayesian inference framework for the lens parameters: there is no clear way to impose explicit priors on them. The fact that rigorous uncertainty quantification is crucial for the science goals enabled by strong lensing suggests that more attention should be given to the exact form in which this inference problem is formulated.

This is fortunately possible, thanks to advances in simulation-based inference methods \cite{2019arXiv191101429C,2019ApJ...886...49B,2020arXiv201007032C}. In this work, by computing the mean of the approximate posteriors provided by Bayesian neural networks and treating them as compressed statistics, we show that it is possible to learn the likelihood function of the compressed statistics given the true lensing parameters from simulations. We calculate the coverage probabilities of sampled posteriors using this likelihood and show that they are accurate while not requiring any tunable calibration process of hyperparameters. This procedure is computationally efficient, allowing the posteriors of hundreds of thousands of lenses to be calculated with minimal computational resources.

\section{Methods}

\paragraph{Simulations}
Strong lensing simulations use the lens equation to relate the coordinates of the image plane $\theta$ to the coordinates of the source plane $\beta$. This relation depends on the scaled deflection angle $\alpha$ through
\begin{equation}
    \beta = \theta - \alpha.
\end{equation}
where $\alpha$ is computed from the mass distribution in the lensing structure.

Our simulations are made of a Singular Isothermal Ellipsoid (SIE) with added external shear for the lensing structure and an elliptical S\'{e}rsic background source. In total, seven parameters are needed to describe the lens and six for the background source. We generate simulations by randomly sampling the lens and source parameters from a uniform distribution, spanning a wide range of lensing configurations. Each simulation contains pixel-wise independent Gaussian noise with standard deviation chosen from a uniform distribution between 1\% and 10\% of the peak lens surface brightness. Additional information on the method used for generating strong lensing simulations can be found in \cite{2017Natur.548..555H}. Examples of our lensing simulations are shown in Fig. \ref{fig1}. 

\paragraph{Data compression}

We train an approximate Bayesian neural network to predict the distribution of the lens and background source parameters from strong lensing images. Approximate BNNs can be used with variational inference to represent the marginalized posterior distribution $p(\theta | x)$ as
\begin{equation}
p(\theta | x) \approx \int p(\theta | x, w) q(w) dw,
\end{equation}
where $q(w)$ is the variational distribution of the network weights. Similar to \cite{2017ApJ...850L...7P}, $q(w)$ is chosen to be a Bernoulli random variable multiplied by the network weights, effectively setting a random number of weights to zero. In practice, this is achieved using dropout \cite{2015arXiv150602142G} on the output of each network layer except the final layer. The distribution over the target lensing parameters $p(\theta | x, w)$ is chosen to be a Gaussian mixture model, expressed as 

\begin{equation}
    \label{gmmeq}
    p(\theta | x, w)  = \sum_k^K \phi_k(x, w) \mathcal{N}(\theta; \hat{\theta}_k(x, w), \Sigma_k(x, w)),
\end{equation}

 where $\phi_k$, $\hat{\theta}_k$ and $\Sigma_k$ represent the weight, mean and covariance matrix for each of the $K$ mixture components. 
 
 Sampling from the approximate marginalized posterior $p(\theta | x)$ is done by first feeding many times the input $x$ to the BNN in order to get different predictions for $p(\theta | x,w)$, where each prediction is made from a different set of weights $w$ due to random dropout. Then, data points are drawn from these predicted distributions, forming a set of samples that cover the distribution $p(\theta | x)$.
 
\label{datacompression}

We train a two-component BNN with $20\%$ dropout rate and another with no dropout, essentially predicting a distribution over $p(\theta | x; w)$ parametrized by a fixed set of network weights. After training, the BNNs are used to get compressed statistics $\hat{\theta}(x)$ of lensing simulations (by computing the mean of predicted posteriors), which, along with the true simulation parameters $\theta$, serve as training data for the mixture density network. The main reason for obtaining our point estimates from these networks (instead of training a simple CNN trained with a Mean Square Error loss, for example) is to be able to compare the results of our likelihood-free inference method to approximate BNN methods for the same network, same training, and same performance.

\paragraph{Density estimation}
\label{dens_sec_2}
Mixture density networks (MDN) are neural networks that model conditional probability densities as a mixture of parametric distributions. Typically, the parametric model is chosen to be a mixture of Gaussian distributions similar to Eq. \ref{gmmeq}. However, an MDN models a conditional probability density parametrized by a fixed set of trained network weights $w$, instead of assuming a variational distribution $q(w)$ over the weights.

The network uses as input $\theta$ and the Gaussian noise standard deviation used in the simulations and outputs a prediction for the parameters $\phi$, $\mu$, and $\Sigma$ to model the distribution over $\hat{\theta}$. Consequently, these parameters are conditioned on the input given to the network, which, in this case, gives us a model for the likelihood. The network architecture is a simple two layer feedforward neural network predicting a two-component mixture model. In order to sample from the posterior, we use this modelled likelihood with the affine-invariant MCMC algorithm implemented in the Python package PYDELFI \cite{2019MNRAS.488.4440A}. The algorithm is written entirely in Tensorflow, allowing for GPU accelerated MCMC sampling.

\begin{figure}
  \centering
  \includegraphics[width=0.9\linewidth, trim = 0cm 0cm 0cm 1cm, clip]{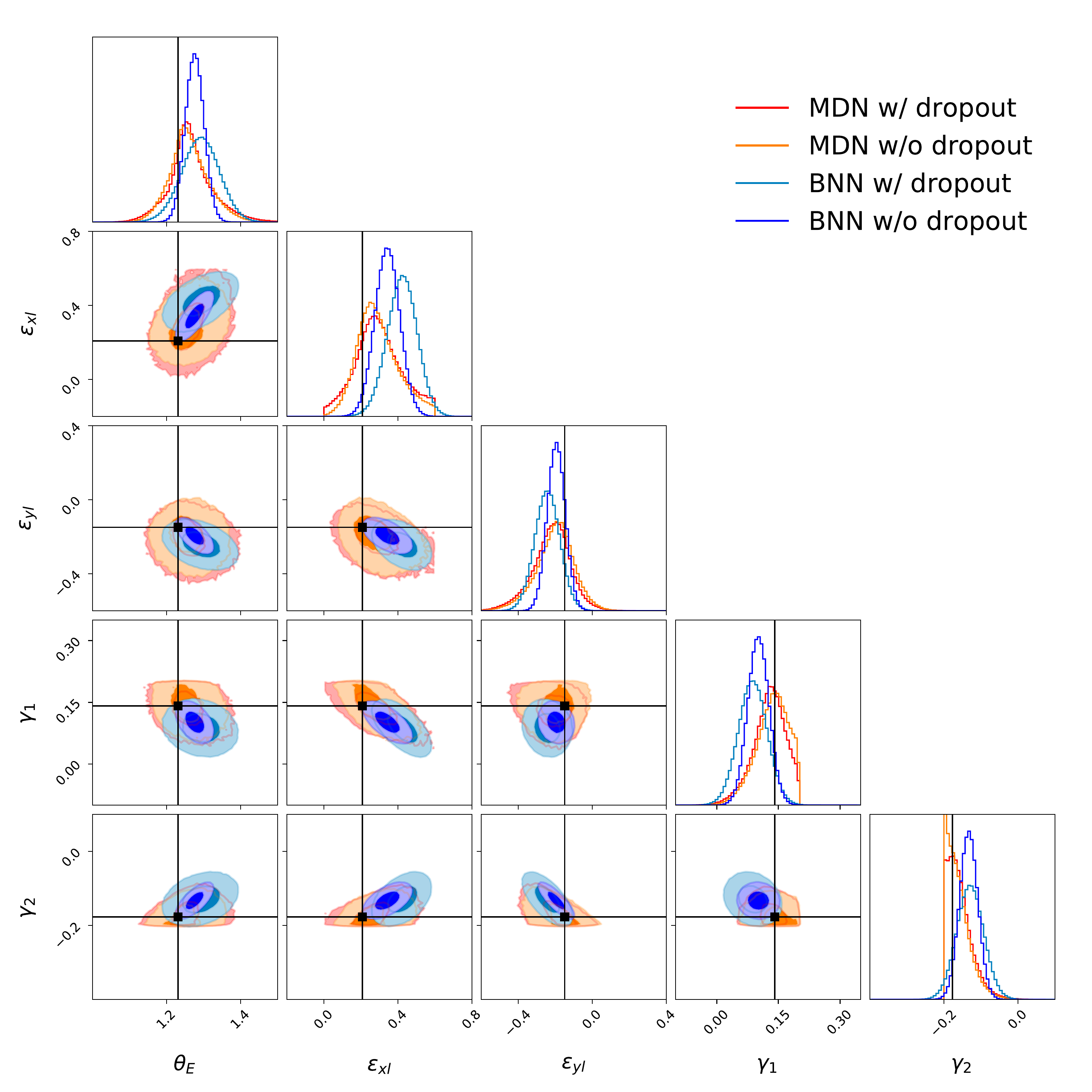}
  \caption{An example of the sampled posterior from a BNN with and without dropout, and from likelihood MDNs trained using compressed statistics of these respective BNNs. Sampling using the MDN likelihood allows for explicitly defining a prior over lens parameters, which in our case is a uniform distribution.}
  \label{fig2}
\end{figure}

\paragraph{Accuracy test}
To evaluate the accuracy of the predicted posteriors, we calculate the coverage probabilities following \cite{2017ApJ...850L...7P, 2021ApJ...909..187W}. Simply put, the test is based on the idea that for an accurate posterior, the true value should be found $x\%$ of the time in a region that encompasses $x\%$ of the total probability.

To perform this test, we generate a set of 10000 lensing simulations and get samples from their predicted posterior distributions. For each simulation, we compute the distance between the true $\theta$ and the predicted lens parameters $\hat{\theta}$ based on a chosen distance metric. Similar to \cite{2021ApJ...909..187W}, we define this distance metric as
\begin{equation}
    d(x) = (x - \hat{\theta})^T \cdot \Sigma_{d} \cdot (x - \hat{\theta}),
\end{equation}
where $\Sigma_{d}$ is the empirical covariance matrix of true lensing parameters over our entire set of simulations.
For every simulation, we compute the fraction of posterior samples that fall within the distance between the true and mean predicted lens parameters. This gives us an estimate of the probability volume needed to include the true value.

We then check over the entire set of simulations the number of times the truth is found within some arbitrary probability volume $x\%$. If the truth falls within this volume $x\%$ of the time, then the predicted posteriors are on average perfectly calibrated. Results of this test are shown in \ref{fig3}.

\section{Results and discussion}
\label{res_sec_3}

\begin{figure}
  \centering
  \includegraphics[width=0.7\linewidth]{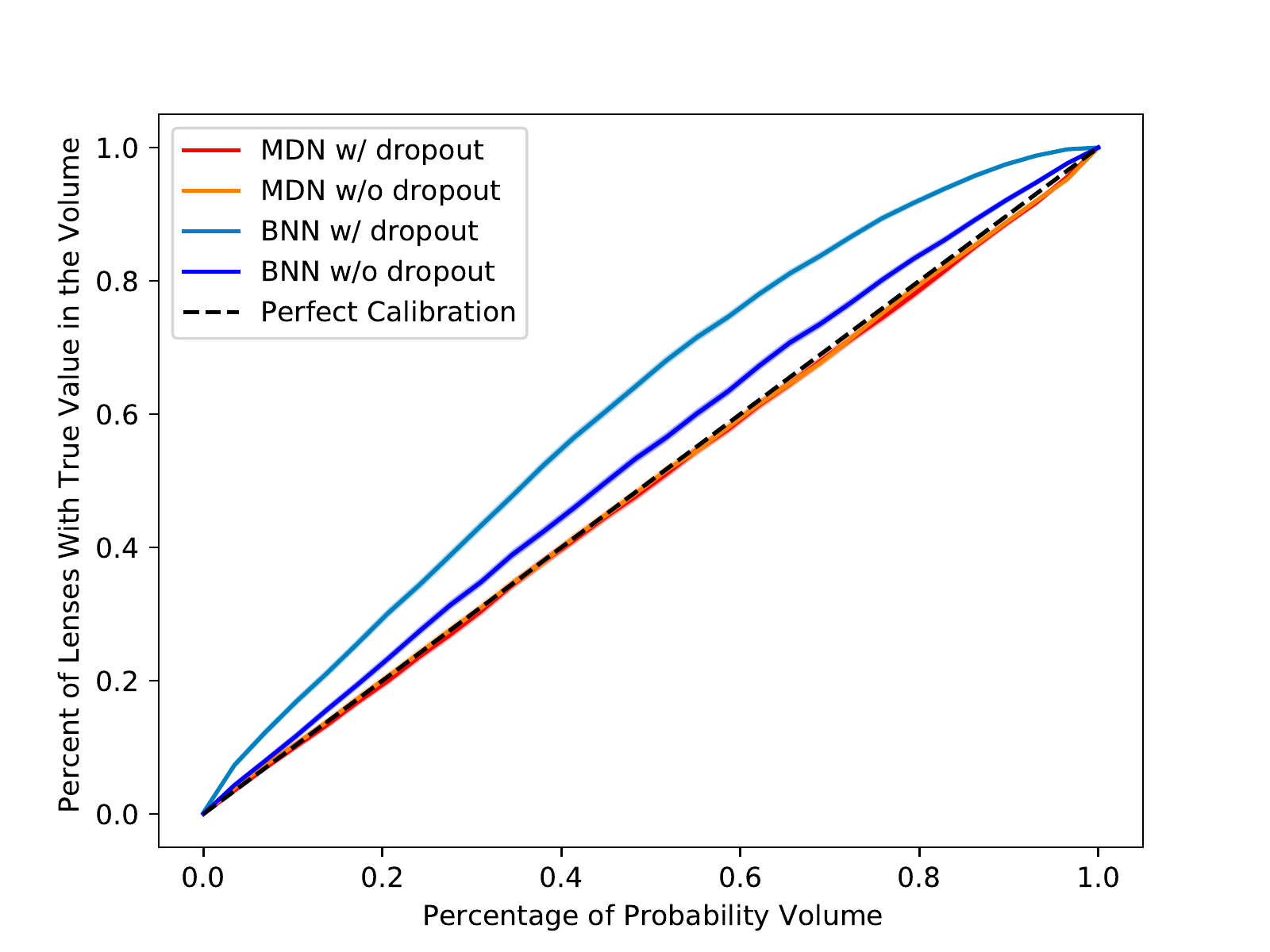}
  \caption{Results from the accuracy test conducted from predicted posteriors over a set of 10000 lensing simulations. For both BNNs, with and without dropout, the predicted posteriors are considered under confident. In the case of the MDN likelihood, this test indicates that the sampled posteriors are on average perfectly calibrated.}
  \label{fig3}
\end{figure}

An example of predicted posterior distributions of lens parameters marginalized over the source parameters is shown in Fig. \ref{fig2}. Although both BNN posteriors appear to be biased, the method using the modelled likelihood remains accurate, despite being trained on compressed statistics from these BNNs. This makes density estimation a powerful alternative for inference: the model learns a direct mapping between $\theta$ and $\hat{\theta}$. Therefore, even if the compressed statistics are heavily biased due to the BNN having inadequate properties (e.g., using a simplistic architecture, badly calibrated dropout rate or under training the network), accurate posteriors can be obtained given that the true distribution is well modelled by the chosen parametric distribution. Fig. \ref{fig3} shows the results of coverage probability tests, indicating near perfect accuracy for the simulation-based method without the need for any manual tuning.

Even if sampling the posterior using an MDN likelihood required running an MCMC, this proved to be extremely fast. On average, we are capable of running MCMC chains for 200 different posterior predictions in parallel, each sampling approximately $10^4$ samples per second using a single NVIDIA V100 GPU. Having $10^5$ posterior samples for each of the 10000 lensing simulations used in Fig. \ref{fig2} took approximately 20 minutes. This shows that LFI density estimation of predicted lensing parameters is well suited for inferring the posterior of tens of thousands of upcoming lenses in a quick and accurate manner.

Although the SIE model has been extremely successful in representing real lensing galaxies \cite{2008ApJ...682..964B}, background sources can have much more complicated morphologies than the elliptical S\'{e}rsic model. This means that the number of parameters needed to describe the source may be too large for MDNs. In future work, we aim to address this issue by training a generative model for the reconstruction of background sources, leveraging the ability to sample many realizations of the source, which we can use to marginalize over it when modelling the distribution of the SIE lens parameters.

It is important to note that the simulations used in this paper did not include foreground effects often present in real strong lensing data such as stellar light from the main lens. Traditionally, these have been often masked out when sampling an explicit closed-form for the posteriors, potentially resulting in biased estimates. In simulation-based inference, these artifacts would simply be added to the simulations used to train the BNN in order to include their contributions in the final predicted posterior. Although the inference of the lensing compressed statistics may be less precise, the method used here should still converge towards accurate posteriors.

In the future, we will further evaluate the performance of our model beyond the accuracy test shown in Fig. \ref{fig3}.  For this, we plan on using the multidimensional Kolmogorov-Smirnov (KS) test proposed in \cite{2015MNRAS.451.2610H}. Generally, the KS test in dimensions higher than one is ill-defined. The method developed in \cite{2015MNRAS.451.2610H} leverages the fact that the total probability mass $\zeta$ contained in a region of highest probability density (HPD) is a unique one-dimensional statistic. Any HPD region is defined by the boundary where the probability values within are greater than at some sample $x$. The key idea is that the values of $\zeta(x)$ for repeated samples $x$ follow a uniform distribution if $x$ is sampled from the probability distribution where the HPD region is taken from. For our purpose, we can use this to verify whether the true lensing parameters can be accurately sampled from the predicted posteriors. This is done by means of a simple one-dimensional KS test which verifies the null hypothesis that $\zeta$ computed using our predicted posteriors follows a uniform distribution.

We would like to emphasize that the purpose of obtaining compressed statistics from a BNN rather than a deterministic CNN is so that we may compare its predicted posteriors with the ones obtained using the density estimation method. However, it is unclear at the moment if using compressed statistics based on samples obtained from BNNs with a non-zero dropout rate could make the density estimation task more difficult: it may be necessary to consider a larger MDN training set or a more expressive likelihood model to better capture the potential added noise that dropout might induce in the compressed statistics. For now, this does not seem to be the case based on Fig. \ref{fig3}: the two MDN models trained on compressed statistics from $0\%$ and $20\%$ dropout BNNs showed on average similar levels of accuracy. The effects of dropout on the MDN-based posteriors will be further explored as we plan on evaluating the performance of our method on additional tests such as the multidimensional KS test described above.

\section{Broader Impact}
This work is contributing to the development of uncertainty estimation methods for the accurate analysis of astronomical data using deep learning. The proposed approach could be applied to other problems in other domains where accurate and detailed quantification of uncertainties is crucial, and its impact could be most important in cases where high-dimensional posteriors need to be characterized.

\begin{ack}
This research was supported by a generous grant from the Schmidt Futures Foundation. The work was also enabled in part by computational resources provided by Calcul Quebec and Compute Canada. Y.H. and L.P. acknowledge support from the National Science and Engineering Council of Canada, the Fonds de recherche du Québec, and the Canada Research Chairs Program. We thank Adam Coogan, postdoctoral researcher at the University of Montreal, for useful discussions in regards to simulation-based inference methods.
\end{ack}

\bibliography{neurips_2021_lfi}

\appendix

\section{Appendix}

\begin{figure}[ht]
  \centering
  \includegraphics[width=1.0\linewidth]{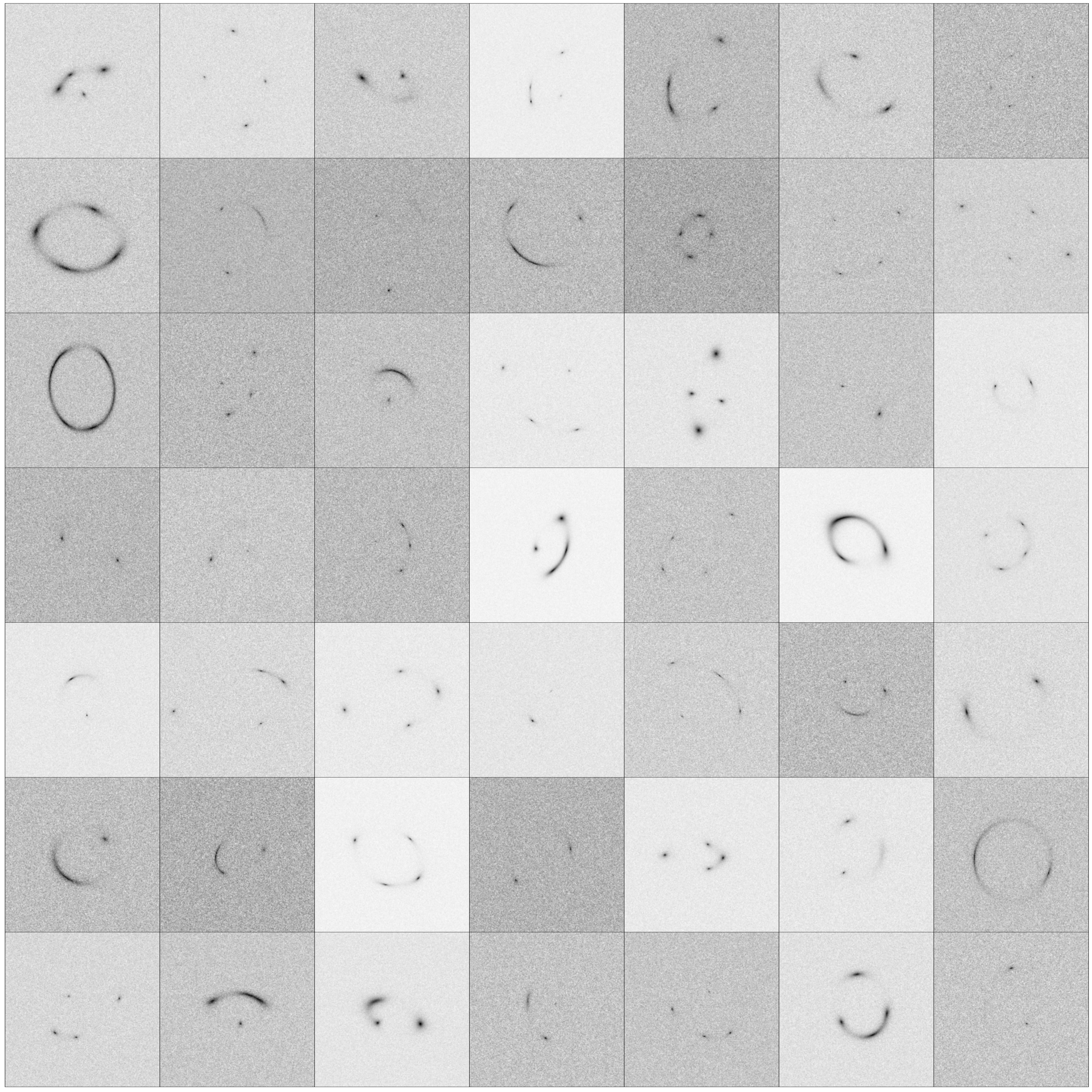}
  \caption{Examples of generated lensing simulations.}
  \label{fig1}
\end{figure}

\end{document}